\documentclass[mathleft]{an}
\usepackage{natbib,twoopt}
\bibpunct{(}{)}{;}{a}{}{,}             
\usepackage{graphicx}
\usepackage{color}
\newcommand{\Teff}{\ensuremath{T_\mathrm{eff}}}
\newcommand{\g}{\ensuremath{g}}

\newcommand{\glog}{\ensuremath{\log\g}}
\newcommand{\beq}{\begin{equation}}
\newcommand{\eeq}{\end{equation}}
\newcommand{\pdx}[2]{\frac{\partial #1}{\partial #2}}

\newcommand{\var}[1]{\ensuremath{\sigma^2_{#1}}}
\newcommand{\sig}[1]{\ensuremath{\sigma_{#1}}}

\newcommand{\lam}{\ensuremath{\lambda}}

\newcommand{\Ai}{\ensuremath{A_i}}
\newcommand{\Bi}{\ensuremath{B_i}}
\newcommand{\Aj}{\ensuremath{A_j}}
\newcommand{\Bj}{\ensuremath{B_j}}

\newcommand{\wa}{\ensuremath{w_\mathrm{a}}}
\newcommand{\wb}{\ensuremath{w_\mathrm{b}}}
\newcommand{\eref}[1]{\mbox{(\ref{#1})}}
\newcommand{\aap}{A\&A}
\newcommand{\nat}{Nature}
\newcommand{\kms}{km\,$\mathrm{s^{-1}}$}

\makeatletter
  \newcommandtwoopt{\citeads}[3][][]{\href{http://adsabs.harvard.edu/abs/#3}%
    {\def\hyper@linkstart##1##2{}%
     \let\hyper@linkend\@empty\citealp[#1][#2]{#3}}}
  \newcommandtwoopt{\citepads}[3][][]{\href{http://adsabs.harvard.edu/abs/#3}%
    {\def\hyper@linkstart##1##2{}%
     \let\hyper@linkend\@empty\citep[#1][#2]{#3}}}
  \newcommandtwoopt{\citetads}[3][][]{\href{http://adsabs.harvard.edu/abs/#3}%
    {\def\hyper@linkstart##1##2{}%
     \let\hyper@linkend\@empty\citet[#1][#2]{#3}}}
  \newcommandtwoopt{\citeyearads}[3][][]%
    {\href{http://adsabs.harvard.edu/abs/#3}
    {\def\hyper@linkstart##1##2{}%
     \let\hyper@linkend\@empty\citeyear[#1][#2]{#3}}}     
\makeatother

\begin{document}

\Pagespan{1}{}
\Yearpublication{2015}%
\Yearsubmission{2015}%
\Month{}%
\Volume{x}%
\Issue{x}%

\title{Stellar science from a blue wavelength range}

\subtitle{A possible design for the blue arm of 4MOST}

\author{ C. J. Hansen\inst{1,2}\fnmsep\thanks{Corresponding author:
  \email{cjhansen@dark-cosmology.dk}\newline}, H.-G. Ludwig\inst{1}, W. Seifert\inst{1}, A. Koch\inst{1}, W. Xu\inst{3}, E. Caffau\inst{4},
 N. Christlieb\inst{1}, A. J. Korn\inst{5}, K. Lind\inst{5}, L. Sbordone\inst{6,7}, G. Ruchti\inst{8}, S. Feltzing\inst{8}, R. S. de Jong\inst{9}, 
 S. Barden\inst{9},  and O. Schnurr\inst{9}}

\titlerunning{Stellar science from a blue wavelength range}
\authorrunning{C. J. Hansen et al.}

\institute{Zentrum f\"ur Astronomie der Universit\"at Heidelberg,
  Landessternwarte, K\"onigstuhl 12, 69117 Heidelberg, Germany \and 
  Dark Cosmology centre, Niels Bohr Institute, University of Copenhagen,
  Juliane Maries Vej 30, 2100, Copenhagen, Denmark \and 
  Optical System Engineering, Kirchenstr. 6, 74937 Spechbach, Germany \and
   GEPI, Observatoire de Paris, CNRS, Universit\'e Paris Diderot, 5 Place Jules
  Janssen, 92190 Meudon, France  \and
   Department of Physics and Astronomy, Uppsala University, Box
  516, SE-75120 Uppsala, Sweden 
  \and
   Millennium Institute of Astrophysics, Chile \and
  Pontificia Universidad Cat\'olica de Chilem Av. Vicu\~na Mackenna 4860, 782-0436 Macul, Santiago, Chile \and Lund Observatory, Department of Astronomy and Theoretical
  Physics, Box 43, SE-22100 Lund, Sweden \and Leibniz-Institut f\"ur Astrophysik Potsdam, An der Sternwarte 16, D-14482 Potsdam, Germany }

\received{10 March 2015}
\accepted{}
\publonline{later}


\keywords{ 
Instrumentation: spectrographs,
Techniques: spectroscopic,
Stars: abundances
}

\abstract{
From stellar spectra, a variety of physical properties of stars can be
derived. In particular, the chemical composition of stellar atmospheres
can be inferred from absorption line analyses.  These provide key
information on large scales, such as the formation of our Galaxy, down
to the small-scale nucleosynthesis processes that take place in stars
and supernovae. By extending the observed wavelength range toward bluer
wavelengths, we optimize such studies to also include critical
absorption lines in metal-poor stars, and allow for studies of
heavy elements ($Z\ge 38$) whose formation processes remain poorly constrained. In
this context, spectrographs optimized for observing blue wavelength ranges are essential, 
since many absorption lines at redder wavelengths are too weak to be
detected in metal-poor stars.  This means that some
elements cannot be studied in the visual--redder regions, and important
scientific tracers and science cases are lost.
The present era of large public surveys will target millions of stars. It is therefore important that the next generation of
spectrographs are designed such that they cover a wide wavelength range
and can observe a large number of stars simultaneously. Only then, we
can gain the full information from stellar spectra, from both metal-poor
to metal-rich ones, that will allow us to understand the aforementioned
formation scenarios in greater detail.
Here we describe the requirements driving the design of the forthcoming survey instrument 4MOST, a multi-object spectrograph commissioned for the ESO VISTA 4m-telescope. While 4MOST is also intended for studies of active galactic nuclei, baryonic acoustic oscillations, weak lensing, cosmological constants, supernovae and other transients, we focus here on high-density, wide-area survey of stars and the science that can be achieved with high-resolution stellar spectroscopy.
Scientific and technical requirements that governed the design are described along with a thorough line blending analysis.
 For the high-resolution spectrograph, we find that a sampling of $\ge2.5$ (pixels per resolving element), spectral resolution of 18\,000 or higher, and a wavelength range covering 393--436\,nm, is the most well-balanced solution for the instrument. A spectrograph with these characteristics will enable accurate abundance analysis ($\pm 0.1$\,dex) in the blue and allow us to confront the outlined scientific questions.
}

\maketitle

\section{Introduction}

Spectra obtained in the near-ultraviolet to blue wavelength range
($300\la\lambda\la 450$\,nm) can carry a wealth of information. In stars,
a blue colour can indicate that the star is hot, metal-poor, or a combination of both. Most metals ($Z\geq 3$) have
their strongest absorption lines in this wavelength region. The heavy
elements ($Z\ge 38$) have their strongest, and sometimes their only
absorption lines accessible from ground-based observatories, blue-wards of
430\,nm \citep{cowan2002, Sneden2003}. This region is typically very
crowded causing many severe line blends. Therefore, this region is often
avoided in spectral analyses, as they require detailed and accurate line
lists to allow for a precise chemical abundance analysis from the
atomic and molecular features. Furthermore, high-resolution spectra are
also needed to better resolve and separate the individual absorption
features of the blending components.

Currently, there are only few high-resolution (i.e., $R = \lambda / \Delta \lambda \ga 20\,000$),
blue-to-near-UV sensitive spectrographs on large ground-based telescopes
($\ge 4$\,m), namely the High-Resolution Echelle Spectrograph (HIRES) at
the 10-m Keck telescope, the High-Dispersion Spectrograph (HDS) at the
Subaru 8.2-m telescope, and the Ulraviolet-to-Visual Echelle
Spectrograph (UVES) at the 8.2-m Very Large Telescope (VLT), all of
which operate down to the atmospheric cut-off at $\sim 300$\,nm.
 The MMT Advanced Echelle Spectrograph
(MAESTRO) at the MMT\footnote{also known as Multiple Mirror Telescope} covers the UV spectral
range down to $\sim315$\,nm, and on the same telescope, the Multi-object
Hectochelle down to 380\,nm. Finally, there are the Magellan Inamori
Kyocera Echelle (MIKE) instrument at the 6.5-m Magellan telescope with a
wavelength limit of 335\,nm, and the High Resolution Spectrograph (HRS)
at the 11-m South African Large Telescope (SALT) for which the bluemost
observable wavelength is $\sim 350$\,nm. 

However, in order to answer some of the important open questions in
astrophysics related to the origin and formation of the Galaxy as well
as the creation of heavy elements (see Sect.~\ref{science}),
it is imperative to design more
blue-sensitive, high-resolution, multi-object spectrographs.  In
particular, in old, metal-poor stars as typically found in the Galactic
halo, we need to study the blue wavelength ranges to detect some of the
heavy elements, as their redder lines are often too weak to be
detectable. To address these questions, we carry out detailed studies on 
when absorption lines are detectable and useful in the blue spectral
range for an abundance analysis.

The 4-metre Multi Object Spectroscopic Telescope (4MOST) will be a new
fibre-fed instrument meant to simultaneously provide high- and
low-resolution spectra. It is planned to be mounted at the VISTA
telescope (Visible and Infrared Survey Telescope for Astronomy) at ESO's Paranal Observatory. Currently, 4MOST is in its
Preliminary Design phase \citep[see][for a detailed overview]{Jong2012,Jong2014}.  It will carry out follow-up and
extended observations for Gaia in addition to the large ongoing Gaia-ESO
Survey \citep[GES;][]{Gerry}.  The high-resolution spectrograph design of 4MOST also includes a green and a red arm, which lead to a wide
spectral coverage (from $\sim 390$\,nm to $\sim 676$\,nm, including significant
gaps; Ruchti et al. 2015, subm.), which allows for studies of substructure in
all major components of the Milky Way (MW), namely the disks, bulge, and
halo.  In addition, the instrument will simultaneously obtain spectra at lower resolution ($R>5000$), and
will pursue both Galactic as well as extragalactic studies. Within the Galaxy, the large number of low- and
high-resolution fibres will provide spectra of up to 2400 stars spread
over a field of at least 4 degrees$^2$ field in one single pointing. Hence, 4MOST
envisions to target up to 30 million objects in its first survey of five year duration. 

Ruchti et al. (2015, subm.) describe how the redder wavelength ranges
($\lambda \ge 450$\,nm) of the 4MOST high-resolution spectrograph were optimized to maximize a versatile
scientific outcome. Here, we focus on the scientific requirements of the high-resolution blue arm that drives the design, how
its wavelength coverage is optimized for chemical tagging, and we describe blending issues that apply to every kind of spectrum densely populated by spectral lines.  

This paper is
arranged as follows: Section~\ref{design} sketches a concept design and
highlights potential construction issues. Section~\ref{science} outlines
the scientific drivers, Sect.~\ref{sec:require} describes the scientific
requirements such as wavelength coverage, resolution, sampling, and
blending issues. Finally, the conclusion and outlook can be found in
Sect.~\ref{conclusion}.


\section{Constructing the blue arm of 4MOST}
\label{design}
\begin{figure}[h!]
\centering 
\includegraphics[width=0.47\textwidth]{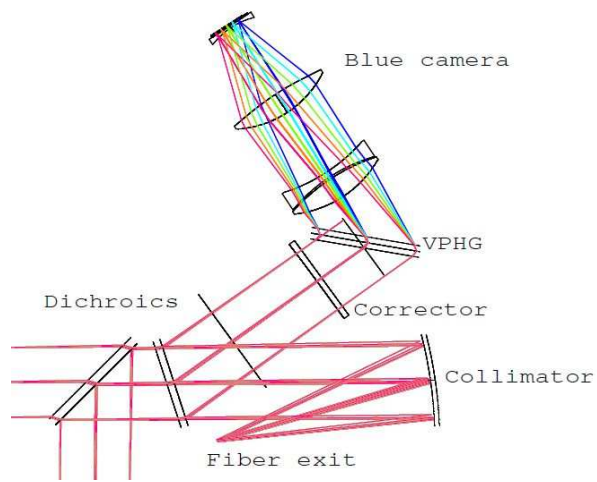}
\caption{Layout of the 4MOST HRS blue channel: common collimator optics,
  separation into three channels by dichroics (here only the blue channel
  is shown), followed by the VPH grating, and finally the camera. 
  \label{fig:design1}}
\end{figure}

The conceptual optical design and layout of the 4MOST High-Resolution
Spectrograph (HRS) is presented in this section. We emphasize that the
details of the design is conceptual, and the information
provided is only intended to guide the reader. The HRS will have three
channels covering the wavelength range from 392.8 to 675.5\,nm; we will
focus here on the blue channel covering the wavelength range from 393 to
436\,nm.

\begin{table}
\begin{center}
\caption{Conceptual design specifications of the 4MOST HRS. \label{tab:layout1}}
\begin{tabular}{l|c}
\hline
\hline
Components$/$Parameters & Values  \\
\hline
Wavelength range & 393-436\,nm (blue)\\
    & 516-572\,nm (green) \\
    & 610-676\,nm (red) \\
\hline
Spectral resolution & $>18\,000$ (goal: 20\,000)\\
\hline
Fiber core diameter & 85\,$\mu$m\\
\hline
Fiber separation & 170\,$\mu$m \\
\hline
Fiber output focal ratio & F3 \\
\hline
Spectrograph beam size & 250\,mm \\
\hline
Collimator & Reversed off-axis \\
       &   Schmidt telescope \\
\hline
Grating & VPHG: \\
        & 3586 l$/$mm (blue)\\
        & 2836 l$/$mm (green)\\
        & 2343 l$/$mm (red)\\
\hline
Camera focal ratio & F1.8 \\
\hline
Detector & 6K x 6K, 15\,$\mu$m$/$pixel \\
\hline
Sampling (spatial direction) & 3.4\,pixels \\
  ~~~~~~~~ (dispersion direction)   & 2.5 -- 3.4\,pixels\\
\hline
Gap between fiber images & 3.4\,pixels \\
\hline
Number of fibres & $\sim$ 840 \\
\hline
\hline
\end{tabular}
\end{center}
\end{table}	 

The main design parameters are given in Table~\ref{tab:layout1} and
the conceptual layout of the blue channel is shown in Fig.~\ref{fig:design1}. Around 840 fibers can be injected in the HRS. After a common collimator
optics, the light is split by dichroic filters into three separate
channels to optimize the efficiency of the
instrument. Volume-phased-holograpic (VPH) gratings are foreseen to
disperse the light in each channel. The camera optics for all channels
is very similar, but not identical, and is also optimized for the
wavelength range to be covered. As detector a CCD with 6\,k$\times$ 6\,k
pixels is under evaluation. With this configuration, a minimum spectral resolving
power of $R=18\,000$ (goal: $20\,000$ -- see also Sect.~\ref{sec:resolution}) is reached over the entire spectral range
covered.  The theoretical optical performance of the system is excellent. 
The mean value of the width that receives 80\% of the energy is 0.7\,pixel over all wavelengths and fields. The poorest value is obtained at the CCD edge and corresponds to 1.5\,pixels. Thus, minimal degradation of the nominal resolution/spectral purity will occur.

\begin{figure}[h!]
\centering 
\includegraphics[width=0.47\textwidth]{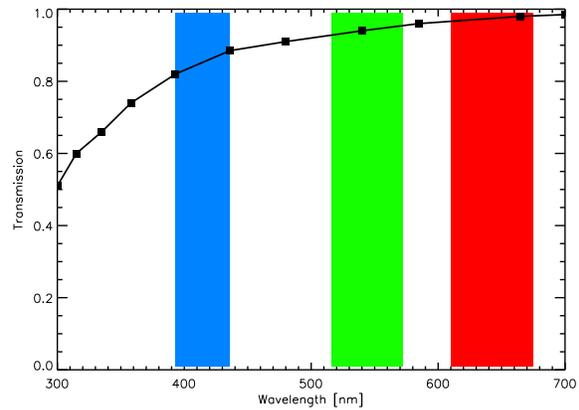}
\caption{Expected throughput for a fibre candidate to be used in 4MOST with a length of
  20\,m. The curve shown is only internal transmission, no coupling losses
  are included. The boxes indicate the wavelength ranges for blue the
  ($\lambda \sim 400$\,nm), green ($\lambda \ge 500$\,nm), and red ($\lambda
  \ge 600$\,nm) channel, respectively.\label{fig:des3}}
\end{figure}

In the blue wavelength range, the efficiency is a
point that needs special attention. The total efficiency of the HRS blue
channel, not including atmosphere and telescope, is given in
Table~\ref{tab:layout2}. For a fiber-fed spectrograph such as 4MOST, the
major limitation comes from the fiber throughput. Therefore,
special measures are taken in the layout of the instrument at the
telescope, such as to minimize the length of the fibers, which for 4MOST
will be about 20\,m. In Fig.~\ref{fig:des3}, the internal transmission of
a 4MOST fibre candidate for a length of 20\,m is shown with the HRS
wavelength ranges indicated. While the throughput is well above 90\,\% for
the green and red ranges, it drops to a mean value of about 85\,\% in the
blue range.

\begin{figure}[h!]
\centering
\includegraphics[width=0.47\textwidth]{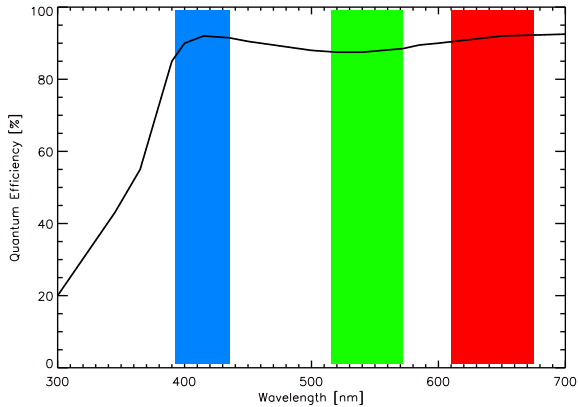}
\caption{Indicative detector efficiency for 4MOST HRS on the basis of a typical
  6\,k\,$\times$\,6\,k chip and a candidate coating. The boxes indicate the wavelength ranges (as in
  Fig. \ref{fig:des3}) for the blue, green, and red channel,
  respectively. \label{fig:design4}}
\end{figure}

A typical, indicative detector quantum efficiency is shown in
Fig.~\ref{fig:design4}.  There are several very efficient coatings for
the detector, giving a high efficiency over the full HRS wavelength
range: a coating candidate is illustrated in
Fig.~\ref{fig:design4}, which in the blue yields an efficiency close to
90\,\%. For the VPH grating, the theoretical efficiency was calculated
and a contingency factor of 0.9 applied, yielding a conservative
estimate. The throughputs of the common collimator, dichroics and camera
are summarized in Table \ref{tab:layout2}. A peak efficiency of 46\,\%
is expected, with 27\,\%$/$30\,\% at the ends of the blue
wavelength range.

\begin{table}
\begin{center}
\caption{Total efficiency of the blue channel not including effects from the atmosphere and telescope. \label{tab:layout2}}
\begin{tabular}{lccc}
\hline
\multicolumn{4}{c}{~~~Blue channel}\\
\hline
Component & 393\,nm & 414\,nm & 436\,nm \\
\hline
Fibers$/$Coupling & 0.74 & 0.77 & 0.79 \\
Collimator & 0.95 & 0.95 & 0.95 \\
Dichroic(s)& 0.98 & 0.98 & 0.98 \\
VPH grating & 0.49 & 0.75 & 0.48 \\
Camera & 0.92 & 0.93 & 0.94 \\
Detector & 0.88 & 0.92 & 0.91 \\
\hline
Total & 0.27 & 0.46 & 0.30 \\
\hline
\hline
\end{tabular}
\end{center}
\end{table}


\section{Science drivers for the blue region}\label{science}

The Milky Way (MW) provides a unique study case for the formation of
barred spiral galaxies. Detailed information on the formation history can only
be accurately extracted from stars in our Galaxy, constraining models of
spiral galaxy formation. The understanding of the involved physical processes
can then be generalized to other galaxies that are too distant to be
studied with the same accuracy. 

Of particular importance in such studies are the chemical abundances of
individual stars. For example, the $\alpha$-elements (such as O, Mg,
Si, Ca, and Ti) provide information on whether or not the star was likely
formed in situ or brought into the Galaxy after a merger event with
dwarf galaxy-like objects \citep[e.g.,][]{Searle1978, Venn2004,
  Koch2009, Nissen2010}. The general low star formation
efficiencies in the low-mass dwarf galaxies and the ensuing paucity of
the $\alpha$-producing Type II Supernovae (SNe) at the time the
Fe-producing SNe Ia start to contribute leads to a low $\alpha$/Fe-ratio
\citep[e.g.,][]{Shetrone2003,Koch2008,Aden2011}, in contrast to the
rapid enrichment the halo experienced creating higher [$\alpha$/Fe] abundances \citep[e.g.,][]{Nissen2010,Ruchti2014}.

At low metallicity ($\la -2$ dex) the low-luminosity dwarf galaxies \citep[e.g.,][]{Gilmore2013,Koch2014} share the same enhancement in $\alpha$-elements as is seen in Galactic halo stars \citep[e.g.,][]{Cayrel2004}. However, this is not the case at higher metallicities, where the stars in the dwarf galaxies tend to show lower $\alpha$-abundances compared to the Galactic stars.
Furthermore, understanding the
 assembly history of our Galaxy also benefits from mapping the dynamics
 of important Galactic substructures through stellar kinematical
 analyses. Thus tracing, e.g., stellar streams from disrupted satellites
 in phase space, ancient disruption events that are still coherent in
 velocity space can be efficiently recovered.  In turn, these accurate
 kinematic tracers, such as streams and stellar velocities, are a
 requirement to constrain the mass distribution of the Milky Way and the
 shape of its dark matter halo (e.g., \citealt{Koposov2010};
 \citealt{Williams2011}).

Altogether, this will also enable a characterization of the progenitor
systems such as their mass, luminosity, star formation rate, and
chemical enrichment histories.  In particular old, metal-poor stars will
provide a better understanding of the early formation and chemical
evolution of the Galaxy. This outlines the importance of accurate
dynamics and abundance studies, for which high-resolution spectrographs
are needed (see also \citealt{Lindegren2013}).  Hence, high-resolution
stellar spectra yielding radial velocities and elemental abundances will
characterize a variety of features from the nuclear formation processes
that created the elements to shed light on large-scale galaxy formation
scenarios.

The stellar halo is the largest (yet least massive) component in the
MW. There is growing evidence from kinematical and chemical abundance
studies that it consists of two distinct subcomponents, an inner and an
outer halo \citep{Carollo2007, Nissen2010},
although these results are still being debated \citep{Schoenrich}. These two components are separated in their
kinematics and elemental abundances, in particular iron abundances, from
which a metallicity distribution function (MDF) can be
calculated\footnote{This MDF agrees with the one obtained
  from low-resolution metallicity indicators such as the Ca triplet or
  the Ca~HK lines.}.  Studies have
shown that the outer halo is characterised by metal-poor stars
($<[$Fe/H$]>\sim -2.2$) on retrograde orbits, indicating that these
stars were accreted from other galaxies, while the inner halo stars ($<[$Fe/H$]>\sim -1.6$) show
prograde orbits, implying that they formed in
situ \citep{Carollo2007}. A deeper, large survey of stars in both the
inner and outer halos is needed to improve our current, incomplete
understanding of hierarchical galaxy formation, and metal-poor stars
play a crucial role for understanding this.  4MOST will consist of low-
($R\sim 5000$ with a goal of $7500)$ and high-resolution $(R\sim
18\,000$; goal $20\,000)$ spectrographs that are optimal for finding
metal-poor stars with $B > 17.5$\,mag, and acquiring high-resolution
spectra of promising metal-poor candidates down to $B = 16$\,mag.

In general, the halo is thought to be the Galactic component containing
the oldest stars, and they are metal-poor. Recently, also metal-poor stars in the direction of the bulge have been found \citep{bulge_paper}, and its old age has been demonstrated in \citet{Clarkson2008}. The spectra of these metal-poor stars show very few detectable metal lines
\citep{christlieb2004,Hansen2014b}, and some of the strongest lines,
still detectable in metal-poor stars, are found in the blue spectral
region. In order to extract the maximum possible information from these
metal-poor halo stars, a blue wavelength range is essential.  

This is
particularly true for the absorption lines of the heavy elements, as
most of these lines are found below $\sim 430$\,nm \citep{cowan2002,
  Sneden2003}.  They provide important tracers of elements created by
the neutron-capture processes. There are two main channels for creating
neutron-capture elements; the $r$apid and $s$low neutron-capture
process.
The exact physics and sites of these processes are poorly constrained, and
this emphasises the need for abundances of heavy elements from blue
high-resolution spectra, to map these processes observationally (see,
e.g., \citealt{Sneden2008,Hansen2011,Hansen2012,Hansen2014a,Hansen2014b}).  

Additionally, the blue range contains
numerous elements useful for chemical abundance analyses of stars with
$\mathrm{[Fe/H]}\leq -1$, but this range can also be used for
chemical analyses of more metal-rich stars. The blue spectral range
hosts not only transitions of neutron-capture elements, but also lines
from elements belonging to each of the following groups: $\alpha$,
odd-Z, (in)complete Si-burning, and Fe-peak elements, making this a
crucial wavelength region for chemical tagging.  Furthermore, the blue
range $\lambda < 450$\,nm contains the $A^2\Delta - X^2\Pi$ G-band of
CH, which can be used to detect carbon-enhanced metal-poor (CEMP) stars\footnote{Metal-poor stars with [C/Fe]$>0.7$ according to \citep{Aoki2007}.},
and to determine their carbon abundances. Combined with the $s$-process
elements (e.g., Sr, Y, Zr, Ba, La) and $r$-process elements (e.g., Eu,
Nd), the blue spectra thus serve to classify CEMP stars into $-s, -r,
-rs$, and $-$no subgroups through their heavy element abundances (e.g., \citealt{Masseron2010,Bisterzo2012}).  This
is important, as many ($20$--$40$\,\% or more;
\citealt{Beers2005,Lucatello2006,Lee2013,Placco2014}) of the extremely and ultra
metal-poor stars are enhanced in carbon, and almost all the ultra Fe-poor stars are enhanced in carbon
\citep{christlieb2004,Norris2007,Frebel2007,Keller2014,THansen,Bonifacio2015}.
  Understanding how these stars form is key to understanding
how the early halo was enhanced in metals, and this in turn will place
constraints on the physics of the formation of the first few generations
of stars in the Universe.

All of the above underlines the need for spectrographs sensitive at blue
wavelengths ($\lambda \la \sim 450$\,nm), and 4MOST would only be the eighth
spectrometre\footnote{Only UVES is accessible to the entire astronomy
  community, while all other high-resolution spectrographs listed above
  are restricted-user access instruments. This would make 4MOST the
  {\emph second} instrument with a blue arm that will be globally
  available.} sensitive in the blue.

4MOST also envisions a low-resolution mode ($R=5000$--$7500$) to cover a
broad range of Galactic science goals. Many of these differ from the
high-resolution science cases, so we will only briefly comment on the
science facilitated by the blue region of the low-resolution setting. 

Even though the blue region can be used for chemical tagging of more
metal-rich disk stars, this region is suboptimal owing to the extreme
line blending in the blue region in metal-rich stars (see
Sect. \ref{blending} - \ref{sec:obsblend} for details on line blending). Therefore, chemical
abundance analyses of metal-rich Population~I stars are generally
conducted at redder wavelengths. Moreover, radial velocities can also be
determined from blue spectra of metal-rich disk stars, but again due to
the large number of heavily blended and unresolvable lines, leading to
more uncertain measurements, the redder regions of the low-resolution
mode of 4MOST, such as the region around the near-infrared Ca triplet
lines, are preferred for this purpose.

An important driver for the blue, low-resolution spectrograph is the
detection and first characterization of hyper metal-poor stars\footnote{Following the definitions in \citet{Beers2005} these stars have $\mathrm{[Fe/H]}<-4$.}
 in the MW halo.  Here, the blue region of choice (goal) comprises
390--543\,nm, to include the Ca~H and K lines (at $\lambda = 397$\,nm
and 393\,nm, respectively) as the
strongest lines that remain detectable at these low
metallicities. The Ba~II 455.4\,nm
line is an important tracer of neutron-capture processes, in
particular of the earliest nucleosynthesis in the Universe, as it remains measurable down to extremely low metallicity in
most stars at the low resolution planned for
4MOST. This line will help us subclassify the CEMP stars and, in turn, place constraints on their formation scenario.
 The red wavelength cut of this region is set by the Mg b triplet and the molecular CN-bands. Depending on the effective temperature and
carbon-enhancement, also the G-band of CH (431\,nm) remains detectable; it can be used for inferring C abundances to a
precision of up to $\pm$0.2 dex.  Note that the Ca~H and K lines can also
serve as a measure of stellar activity in young stars.

%
\section{Scientific requirements to the blue instrument \label{sec:require}}
\subsection{Sampling and resolution}	

The spectral resolving power, $R=\lambda/\Delta\lambda$, and sampling
$s$ (pixels per resolving element) are amongst the most important quantities that drive the design of a
spectrograph, and the science that can be achieved. In order to
determine optimal values, taking into account technical as well as
financial boundary conditions, we followed an approach similar to that
outlined in \citet{Caffau2013}. 
In the following we will only give a brief summary of our procedure, while our detailed tests will be presented in a forthcoming paper.

\subsubsection{Synthetic lines and parameter space\label{sec:resolution}}

First, we synthesized a Gaussian line profile, centered at $\lambda_0$,
of given equivalent width (EW) and Full Width at Half Maximum (FWHM) at
initially very high sampling over a continuum level of unity.
The FWHM is chosen to be the resolution element (RE) of the
spectrograph, so that the resolving power of the simulated spectrum can
then be expressed as $R=\lambda_0$\,/\,RE\,\,=\,$\lambda_0$\,/\,FWHM
[nm].  Next, the sampling of the simulated spectrum is degraded to a
varying number of pixels per RE, and Poisson noise is injected to mimic
different signal-to-noise ratios ($S/N$).
The goal is to find a minimum set of requirements, that will still allow us to derive 
abundances with an uncertainty of $\pm 0.1$\, dex or better.
This corresponds to a relative EW uncertainty of 0.22--0.25\footnote{for weak lines and less for strong or partly saturated lines.}.

For testing purposes, we ran 10\,000 Monte Carlo realizations with a fixed S/N of the 256
different combinations of each of the parameters in steps of: EW\,=\,(1.25, 2.5, 5, 10) pm; $R$\,=\,(15\,000 -- 23\,000); $s$\,=\,(1 -- 4) pixels/RE; and $S/N$\,=\,(50, 60, 75, and 100 per pixel, or 154, 185, 231, 308 per \AA\footnote{we used $S/N_{pixel} \propto S/N_A  \sqrt{\frac{\lambda}{R \cdot s}}$  and we calculated the $S/N$ per \AA~ for this specific example with a fixed value of $s=2.1, R=19\,000$, and $\lambda=4200$\AA\,.}). This test was performed to explore the parameter space, and following find a set of useful parameters to explore in more detail.
In each of the Monte Carlo trials, the position of the
resampled pixels was rigidly shifted by a random (linear) amount of up
to $\pm0.5$ times a resampled pixel with respect to the default
value. This is to ensure that the position of the line center is
different each time with respect to the positions of the resampled
pixel, so as to avoid biases due to constant line-pixel
offsets.

\subsubsection{Line measurement, optimal resolution and sampling}

The resampled and noise injected lines were evaluated by fitting a
Gaussian line profile within $\pm3\times$RE around the nominal line
center. The continuum level, in turn, was estimated in two windows of $5\times$REs width and located at $\pm4\times$RE from the line center.

The first tests showed that a sampling of 1 pixel/RE is simply too low to retrieve 
abundances within the required accuracy, and we find that the
sampling needs to be between the lower limit of 2 (Nyquist sampling) and 3\,pixel/RE. 

\begin{figure}[!htb]
  \begin{center}
   \includegraphics[width=1\hsize]{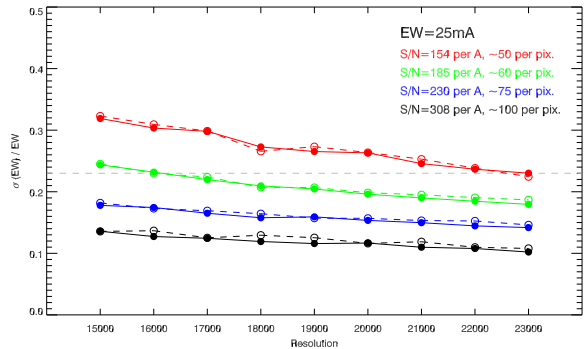}
  \end{center}
  \caption{Relative EW uncertainty vs. resolution for different rates of S/N
    ratio, and a sampling of 2 and 3\,pixel/RE for the filled and open circles, respectively. An indicative relative EW uncertainty is plotted as a dashed, grey line.}
  \label{resolutionsampling}
\end{figure}

After exploring the parameter space, a more detailed test was carried out on the constrained set of promising parameters using finer steps compared to the aforementioned first test. Figure~\ref{resolutionsampling} shows our results (using the restricted sampling of
$s=2$ and $3$\,pixel/RE) in terms of the
relative EW measurement uncertainty, $\sigma$EW/EW, as a function of the
tested resolution.  
A line of 25\,m\AA ~ was chosen to mimic a weak line such as the 670.6\,nm Li line. 

Imposing our requirement that lines need to be measured to a precision of
$\pm 0.1$\,dex elemental abundance to perform
meaningful science for Galactic archaeology, and that the S/N should be obtainable 
within a reasonable exposure time (e.g., S/N=170 per \AA~ or 55 per pixel), a trade off must
be made. In conjunction with the preliminary instrument design, we chose the value of 2.5\,pixel/RE as a compromise\footnote{We set a goal of 3\,pixel/RE instead of the Nyqist-optimal of 2\,pixel/RE to maintain resilience against CCD cosmetics defects that may fall over a measured line. Dispersion variations along the spectral range can make sampling locally worse, so here we assume 2.5\,pixel/RE as a compromise.}.
This leads to our final specifications, namely:
\begin{itemize}
\item $R>18\,000$ everywhere in the covered wavelength range; 
\item a sampling of at least 2.5\,pixel/RE (goal: 3.0\,pixels/RE).
\end{itemize}

The adopted (goal) abundance uncertainty of 0.1\,dex is a value that, provided we have accurate atomic physics input, is well documented. It arises as a typical uncertainty owing to the influence from the stellar parameter uncertainty. This has been tested in great detail for 18 elements \citep[see Table 6 and 7 in][]{Cayrel2004}. In Sect.~\ref{sec:obsblend} we will investigate how this uncertainty affects the usefulness of blends.

\subsection{Selecting and optimizing the blue wavelength range}\label{wavelength}

In order to fulfil the science requirements for the 4MOST high-resolution survey focussing on the halo,
elements from each major group ($\alpha$-, odd-Z, iron-peak, and
neutron-capture elements) should be included in the covered wavelength
range. Such elements can be efficiently measured from absorption lines
in the blue range of metal-poor halo stars, whereas in more metal-rich
stars from the disks and bulge, a redder wavelength region is preferred
owing to fewer line blends (Ruchti et al. 2015, subm.). As part of the
4MOST requirements, we want to be able to derive accurate abundances
($0.1$\,dex uncertainty) from several elements belonging to each
group. Here we test the 390--460\,nm (goal) region and require that each absorption line reaches 0.9
in residual flux, to be detectable even in noisy spectra.

\begin{figure}[h!]
  \centering
    \includegraphics[scale=0.42]{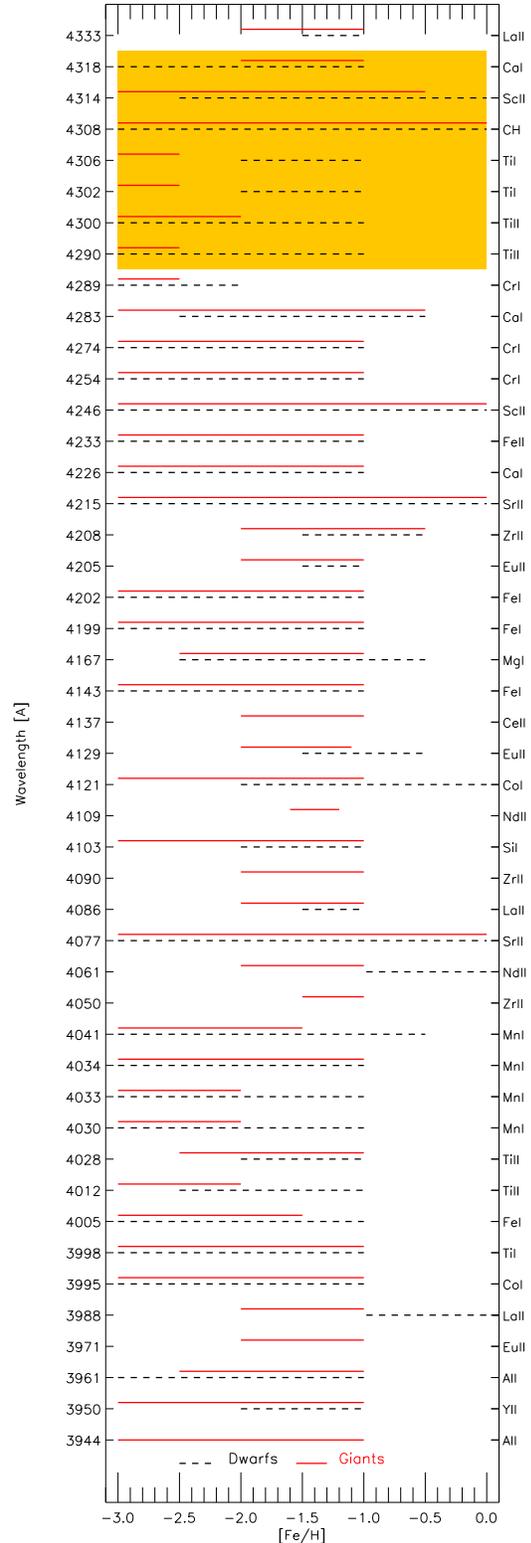}
    \vspace{-0.2cm}
  \caption{Detectable lines in the blue wavelength range in dwarfs
    (5500\,K/4.0/-1,-2,-3/2.0\,km\,s$^{-1}$ - dashed, black line) and giants
    (4500\,K/2.0/-1,-2,-3/2.0\,km\,s$^{-1}$ - solid, red line). Only a few Fe lines
    have been shown for simplicity. The yellow region indicates lines that could be lost, if the star has a
    strong G-band.}
  \label{lines}
\end{figure} 

Synthetic spectra were created with the MOOG synthetic spectrum code \citep[][version
  2010]{Sneden1973}, in conjunction with the MARCS stellar atmosphere models
\citep{Gustafsson2008} matching the stellar parameters listed below. A
line list downloaded from the VALD database \citep{VALD2000} was used for this
purpose. Several transitions in the line list were checked and updated
with the atomic data given in NIST, and \citet{cowan2002}, \citet{Sneden2003}, and 
\citet{McWilliam1995}. The two first mentioned studies focus on heavy elements,
while \citet{McWilliam1995} provides lists of both light and heavy elements and
shows the detectability of numerous elements observed in various stars
at a resolution matching that of the 4MOST HRS. The goal is to obtain a
set of useful, detectable absorption lines of: Mg, Al, Si, Ca, Mn, Fe,
Co, Ni, Sr, Y, Zr, Ba, and Eu. To probe the detectability, we inspected
synthetic spectra in which all abundances are scaled to Solar (i.e.,
$\mathrm{[element/Fe]} = 0.0$). We target lines that are as clean as
possible, reaching the aforementioned 0.9 in residual flux, and we avoid
saturated lines, to ensure that the lines are suited for chemical
abundance analyses. This means that we need at least a $S/N$ of about
$30$ (per pixel) to ensure 3$\sigma$ detections.  In Fig.~\ref{lines}, we show the
detectability in a typical dwarf and giant star as a function of
metallicity, where the stellar parameters ($T_{\mathrm{\scriptsize eff}}$/$\log g$/[Fe/H]/$V_t =$ 5500/4.0/$-1,-2,-3$/2.0\,\kms, and
$4500/2.0/-1$,\\$-2,-3$/2.0\,\kms, respectively) have been tested. These stellar parameters are 
consistent with the study of \citet{Caffau2013}.

Furthermore, also the radial velocity needs to be known to better than
$1-2$\,\kms, a value that was set to be able to accurately trace
kinematic substructures such as streams from disrupted satellites in
phase space. In this way, ancient disruption events that are still
coherent in velocity space can be efficiently recovered.  
Finally, as we will see in Sect.~\ref{sec:obsblend}
an accurate radial velocity is also important to recover abundances for
blended lines.
 
For finding the optimal wavelength range, a trade off between an as
large as possible number of included lines on one hand, and fibre
efficiency and the costs of the blue arm of the instrument on the other
hand, has to be considered. Taking these constraints into account, we
arrived at the wavelength region 393--436\,nm (after scanning the entire goal region: 390-460\,nm). 
This range ensures that
at least one element of each element group is detectable, and can be
measured in either dwarfs, giants, or both, at the required 4MOST
resolving power of $R>18\,000$.

\begin{table}
\begin{center}
\caption{List of lines (elements, Z) tested for detectability in dwarfs/giants at [Fe/H]=$-1,-2,-3$. The `worst' losses are indicated with `*'. Columns three and four fall outside the covered wavelength range. \label{tab:online}}
\begin{tabular}{cl cl}
\hline
\hline
Wavelength [nm] & Z &  Wavelength [nm] & Z \\
\hline       
394.4 & Al I &   439.8 & Y II \\   
395.0 & Y II &   442.9 & La II \\  
396.1 & Al I &   443.5 & Eu II \\  
397.1 & Eu II&   444.3 & Ti II \\  
398.8 & La II&   444.6 & Nd II \\  
399.5 & Co I &   444.9 & Dy II \\  
399.5 & La II&   445.4 & Ca I \\   
399.8 & Ti I &   445.9 & Ni I*\\   
399.9 & Ce II&   446.1 & Fe I \\   
400.5 & Fe I &   446.2 & Nd II \\  
400.5 & V I  &   446.8 & Ti II \\  
401.2 & Ti II &  448.2 & Fe I  \\  
401.9 & Th II &  449.4 & Fe I  \\  
402.1 & Nd II &  450.1 & Ti II \\  
402.3 & Nd II &  453.4 & Ti II  \\ 
402.8 & Ti II &  455.4 & Ba II*\\  
403.0 & Mn I  &  456.3 & Ti II  \\  
403.3 & Mn I   	 && \\
403.4 & Mn I   	 && \\	   		      
405.0 & Zr II    && \\
405.7 & Pb II    && \\
406.1 & Nd II    && \\
407.7 & Sr II    && \\
407.7 & Dy II  &&\\
408.6 & Th II  &&\\
408.6 & La II   && \\
409.0 & Zr II   && \\
410.9 & Nd II   && \\
412.1 & Co I   && \\
412.3 & La II   && \\
412.9 & Ba II   && \\
412.9 & Eu II   && \\
413.7 & Ce II   && \\
414.3 & Fe I   && \\
416.7 & Mg I   && \\
419.9 & Fe I   && \\
420.2 & Fe I   && \\
420.5 & Eu II   && \\
420.8 & Zr II   && \\
421.5 & Sr II   && \\
422.6 & Ca I   && \\
423.3 & Fe II   && \\ 
424.6 & Sc II   && \\ 
425.4 & Cr I   && \\
427.4 & Cr I   && \\
428.9 & Cr I   && \\
429.0 & Ti II   && \\
430.0 & Ti II   && \\
430.2 & Ti I   && \\
430.6 & Ti I   && \\
430.8 & CH   && \\
431.7 & Zr II   && \\
431.8 & Ca I   && \\
431.8 & Sm II   && \\ 
432.2 & La II   && \\
 432.9 & Sm II &&\\
 433.3 & La II && \\
\hline
\hline
\end{tabular}
\end{center}
\end{table}

A few elements (Ni, and Ba) cannot be measured at this resolution in this
range\footnote{That is, unless the star is enhanced in either of the
  elements or if the range is extended to $\sim460$\,nm (goal region).}, and we need to use
lines from the green and red wavelength ranges (Ruchti et al. 2015, subm.). However,
in addition to the elements listed above, the blue range also offers a
large number of other elements such as C (CH), Sc, Ti, Cr, La, Nd, Ce (illustrated in Fig.~\ref{lines}). 

With a considerable number of both lighter and heavy elements well
represented in the blue wavelength range, we can use the 4MOST spectra
to accurately map the fossil gas in old stars. Our optimized range will
also provide a considerable number of Fe and Ti I and II lines (minimum
7 and 5, respectively, covering excitation potentials of $\sim
1-3$\,eV), needed for determining effective temperatures via excitation
equilibrium, and surface gravities via ionisation equilibrium. Only the
cleanest Ti and Fe lines are shown in Fig.~\ref{lines} for the sake of
clarity. This means that the blue region can, on its own, provide
sufficient information needed for a full spectrum analysis. This is
especially the case in metal-poor stars at $\mathrm{[Fe/H]} \la -1$,
and the total number of clean lines will increase when including the
green and red wavelength ranges, as envisioned for 4MOST.

The HRS blue wavelength range was also set such that key elements like
Al, Y, Sr, and CH (plus continuum redward of the band head), which are
located at the edges of the wavelength range, would remain inside the
observed range even for high radial velocity stars which are common in
the Milky Way halo. A wavelength piece of $\sim20$\,{\AA} is required in
the beginning and end of the wavelength range before/after the first/last
line of interest. This allows for loosing 5--10\,{\AA} in the beginning
of the spectrum to noise and even in stars with radial velocities of
$\ga1000$\,km\,s$^{-1}$, we will still be able to get clean detections
of Al in their stellar spectra.

The lines checked in the blue range can be found in Table
\ref{tab:online}. Several of the weak lines may be detectable or useful
if the star is enhanced in the element, or if the $S/N$ ratio is larger
than 30--50 per pixel. Not all the weak lines have been listed, as the line list 
contains more than 4000 lines, most of which are too weak to be detected in 
several of the tested stellar spectra at this resolution and $S/N$ ratio.

\subsection{Blending issues}

The blue part of the spectrum is a very crowded region, and as such, most
absorption lines blend into each other. 
To investigate how strongly blended two lines can be, and still be useful
for abundance determination, we start by investigating the idealized
problem  of fitting two lines with Gaussian profiles, $G(\lambda) \sim a \cdot \exp\left(-\frac{1}{2}\left(\frac{\lambda - \lambda_0}{w}\right)^2\right)$. The goal
is to obtain a relation between the line separation and the EW uncertainty
resulting from the mutual blending.
Following, we test this theory using synthetic spectra mimicking
those from the future 4MOST HRS.

\subsubsection{A simple model of a two line blend \label{blending}}

In this and the following subsection, we consider two weak lines (a, b) with Gaussian line profiles, and we assume that there are no other blending lines interfering with our tests. Furthermore, we assume that the stellar parameters, atomic data, and radial velocity are perfectly constrained, so that the line centres are at the rest wavelength. We want to find the minimum line separation needed to fully recover the abundances of two blended lines.

We consider two lines in a normalized spectrum characterized by two profile
functions $A\ge 0$ and $B\ge 0$, scaled by amplitudes $a$ and $b$, respectively. We
assume that the spectrum is sampled at $m$ points, and that the noise in each
pixel is independently normally distributed with inverse S/N ratio
$n$. We represent the lines in ``depression format'', i.e. fitting the flux
difference to the continuum. The data model $F$ of the spectrum is in this case
\beq
F(\lam) = a A(\lam) + b B(\lam) 
\label{e:datamodel}
\eeq
where \lam\ is the wavelength. We want to determine the amplitudes ($a$, $b$)
scaling the line profiles of the observed data to be fitted $f_i$. Later, we specifically assume Gaussian
profile functions $(A,B)$ of given widths $(w_a,w_b)$ at wavelength positions
$(\lam_a, \lam_b)$.  
We determine $a$ and $b$ by maximum likelihood (hereafter ML) estimation. The problem
is linear in $a$ and $b$ and we obtain as condition for $a$ at the maximum of
the (log) likelihood~$L$  
\beq
\pdx{L}{a} = \sum_{i=1}^m \left(\frac{f_i-F_i}{n^2}\right) \Ai  = 0\enspace,
\label{e:line2dlda}
\eeq 
with the analogous relation for $b$.

We determine the uncertainties of $a$ and $b$ in the standard way from the
inverse of the Fisher information matrix of the problem and obtain
\beq
\var{a} \ge 2 \cdot n^2 \frac{\sum_i \Bi^2}{\sum_{i,j} (\Ai\Bj-\Bi\Aj)^2}\enspace,
\label{e:vara}
\eeq
and
\beq
\var{b} \ge 2 \cdot n^2 \frac{\sum_i \Ai^2}{\sum_{i,j} (\Ai\Bj-\Bi\Aj)^2}\enspace.
\label{e:varb}
\eeq
As expected, the standard deviations scale inversely proportional to the S/N.
Strictly speaking, relations~\ref{e:vara} and~\ref{e:varb} give only lower (the
Cram{\'e}r-Rao) bounds of the variances of $a$ and $b$. However, ML estimators
are efficient and usually get close to the Cram{\'e}r-Rao bound, so that we consider
the lower bounds as actual uncertainties. 

We are interested in how the S/N ratio influences the accuracy to
which the individual amplitudes $a$ and $b$ can be measured to ultimately
obtain an estimate of the associated uncertainties in the derived elemental
abundances.  The problem of determining the EW of a blended line
can be separated into determining the isolated clean line's uncertainty
stemming from noise, and finding a geometrical factor describing the blending
configuration. When fitting a \textit{single line in isolation} our simple model gives
an uncertainty of the amplitude $a'$ (assuming a profile function $A$)
\beq
\var{a'} \ge n^2\frac{1}{\sum_i\Ai^2}\enspace.
\label{e:varaiso}
\eeq 

To have simple analytic expressions at hand, we now approximate the sums in
the variances by integrals assuming Gaussian line profiles of a given,
unique width ($\wa=\wb\equiv w$). For that, we have to assume a sufficiently
fine sampling of the line profile, similarly to what was done in the
derivation of Cayrel's formula \citep{Cayrel1988}. We obtain for the
uncertainties \var{a} and \var{a'} the relation
\begin{eqnarray}
\var{a} & \approx & 
\var{a'} \left(1-e^{-\frac{1}{2}\left[\frac{\lambda_a-\lambda_b}{w}\right]^2}\right)^{-1} \nonumber\\
\mbox{} & \approx &
\var{a'} \left(1-e^{-2.77 \times
  \left[\frac{\lambda_a-\lambda_b}{\mathrm{FWHM}}\right]^2}\right)^{-1} \enspace.
\label{e:vara3}
\end{eqnarray}

Interestingly, equation~\eref{e:vara3} states that the detrimental
effect of blending is independent of the strength of the blending line as long as it remains on the linear part of the curve of growth. This is in agreement with the findings in \citet{Cayrel2004}, where the $\sigma_{\mathrm{EW}}$ is independent of EW. 
 A penalty is already incurred when trying to fit two lines when there 
 is in fact only one.

For our Gaussian line profiles, the equivalent width, $EW$, is given by $EW=\sqrt{2\pi}
a w$, so that error propagation gives $\var{EW}=2\pi w^2 \var{a}$. This
means that equation~\eref{e:vara3} can also be interpreted as a relation
between the uncertainties of the EW of a blended line to the
corresponding value in isolation.

\begin{figure}[h!]
\centering
\resizebox{\hsize}{!}{\includegraphics{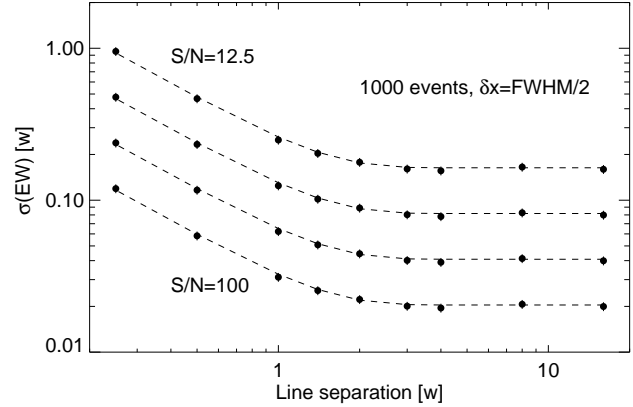}}
\caption{Uncertainty $\sig{EW}$ of one component in a blend of two
  Gaussian components as a function of their separation (both given in
  units of $w$). The analytical result according Eqs.~\eref{e:vara3} is
  depicted by the dashed lines, results of a Monte-Carlo simulation by
  solid dots. $\pm 3\sigma$ error bars (so small that they are barely visible) indicate the
  uncertainty of the Monte-Carlo result. Four cases with S/N=12.5, 25,
  50, and 100 per pixel are shown from top to bottom.
  \label{f:blendtwogauss}} 
\end{figure}

We checked our result by Monte-Carlo (MC)
simulation using two weak lines of the same strength with $a=b=0.1$. Figure~\ref{f:blendtwogauss} illustrates the result of 1000
noise realizations. The MC test confirms that the $\sigma_{\mathrm{EW}}$ is independent of EW (and therefore of line strength) for weak lines. For the MC run we chose a rather coarse (step size
$\Delta\lam=\mathrm{FWHM}/2$) sampling, to show that the sampling is not
particularly important for the validity of the analytical result -- as
long as it is not overly coarse. We note that the way we set-up the MC
run leads to correlated noise among the various S/N cases. This explains
that shifts relative to the analytical result appear the same for all
S/N. All events for a particular S/N are independent. To double
$\sig{EW}$ in comparison to the case in isolation one needs a blending
line as close as 
\begin{eqnarray}
\Delta\lam_\mathrm{double}&=&\sqrt{2\ln(4/3)}\,w\approx 0.759\cdot w \nonumber\\
&\approx &0.322\cdot \mathrm{FWHM} 
\end{eqnarray}
At this separation two Gaussians appear essentially like a single one.
If we instead look at the separation needed in order to derive abundances with an accuracy of 0.1\,dex or better, the necessary separation is $\sim0.6\times$FWHM. This value should be compared to the tests conducted and described in the next section.

It might appear surprising that blended components can be separated to a distance
of $\approx 3/5 \times$FWHM according to the simple model presented
in this section. However, the key property which makes this possible is the
a-priori knowledge of the width of an isolated line. Close blends can be
identified by a certain degree of additional broadening beyond what is
expected for an isolated line. Conversely, the separation becomes more
difficult when the intrinsic line width is not, or only approximately, known.

In the next section we test this in practice, but we also change perspective and address the question of what
happens if uncertainties in the wavelength calibration or measured radial
velocity lead to changes of the position of a blend with respect to the
expected position based on the laboratory rest wavelengths of the blending
components. 

\subsubsection{Probing the line-to-line separation needed for an accurate
 chemical analysis of blended lines
\label{sec:obsblend}}

Following the same assumptions as outlined in the previous subsection we test the needed line-to-line separation to recover the stellar abundances in practice. For this purpose we use synthetic spectra (mock data) that mimic those to be observed by 4MOST in the HRS setting. The two lines are weak, yet not so weak that they mistakenly could be taken as noise. The only free parameters are the two abundances, which we try to recover from spectra synthesized with different line-to-line separations (in units of FWHM) of the two blending lines. 
The lines are treated in isolation and analysed without any `a priori' (input) information from other lines of the same species.

To test how close two blended lines can be, and remain useful for an
abundance analysis, allowing for uncertainties of $0.1$\,dex, we
start by selecting a region where this can be tested in the cleanest
possible way.  

In the bluest part of the blue range, this will be hard
to test, since many blends will arise simultaneously, especially for the
more metal-rich stars. Therefore, we choose a line that was clean in all
the tests made for setting the wavelength region (see
Sect.~\ref{wavelength}). The Cr I line at 427.48\,nm provides a
good test case, as it is clean in both dwarfs and giants from [Fe/H] =
$-3$ to $-1$\,dex. We then insert a line (here we chose Al as an example of a blend) in the core of the Cr line and in small steps move the centre of the Al line away from the centre of the Cr line. We generate synthetic spectra with separations in steps of 0.1$\times$FWHM\footnote{i.e., finer steps were used, compared to the step size in previous section, to probe the separation needed.}. The atomic data, adopted from a nearby Fe I line, was slightly
altered to make this `Al' line as strong\footnote{i.e., reach the same residual intensity} as the Cr line and visible at all
three metallicities ($-3, -2, -1$) in both dwarfs and
giants.
To generalise this test to other elements that only show one blended line within the covered spectral range, we do not use abundance information from lines of the same species at other wavelengths. There are several other useful Cr and Al lines in the covered spectral range, but this is not the case for some of the other (heavier) elements. In this way we have two blending lines for which we have no a priori abundance information from other clean lines.

The test was carried out using noise-free spectra generated with the spectrum synthesis code used in Sect.~\ref{wavelength}, where the lines were broadened to the aforementioned instrumental resolution. Only the abundances of the two blending lines were kept as free parameters and the central wavelengths of these lines are assumed to be known.
This allows us to simultaneously fit both lines and see if the abundances can be recovered with uncertainties better than
$\pm0.1$\,dex. 

\begin{figure}[h!]
\centering
\resizebox{\hsize}{!}{\includegraphics{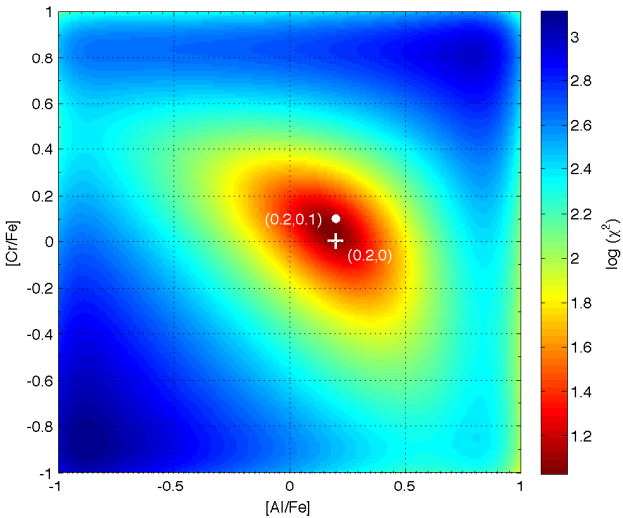}}
\caption{Reduced $\chi^2$ for the automated simultaneous fit to the isolated, blended lines of Al and Cr with a 0.3$\times$FWHM separation. Their abundances are shown on the x- and y-axis, respectively. Contours are smoothed with a kernel of 0.1\,dex width, and the '+' marks the best solution (0.2,0.0) vs the true value 'dot' (0.2,0.1). \label{fig:chi2}} 
\end{figure}

We set up a suite of 10 synthetic spectra with [Cr/Fe]$=0.1$ and [Al/Fe]$=0.2$\,dex and with different line-to-line separation. These spectra were blindly analysed in three different ways by several co-authors.
First we conduct an MC test on a randomly selected spectrum of a metal-poor dwarf star (\Teff/\glog/[Fe/H]/Vt: 5500K/4.0\,dex/$-2.0$\,dex/2\,\kms).

1) We create a grid of synthetic spectra with Al and Cr abundances varying in steps of 0.1\,dex (the adopted allowed uncertainty) and derive the abundances using $\chi^2$ minimization.
For this purpose we adopt a nominal uncertainty of 3\% for the $\chi^2$-fitting i.e., the uncertainty on the data to be fitted, which roughly corresponds to an overall 0.1\,dex abundance uncertainty. At a separation of 0.3$\times$FWHM we recover the Al abundance ([Al/Fe]$=0.2$\,dex) exactly, while we obtain a Cr abundance lower than the input, at [Cr/Fe]$=0.0$\,dex, which is at the limit of the allowed uncertainty. This case is shown in Fig.~\ref{fig:chi2} for which we find a reduced $\chi^2$ of 0.93.

2) A manual test was made on the same spectra by blindly analysing them in MOOG and treating them as real observed spectra for the same dwarf star (5500K/4.0\,dex/$-2.0$\,dex/2\,\kms). At a separation of $0\times$FWHM the input Cr and Al abundances could not be recovered, but we find several possible solutions. At a separation of $0.5\times$FWHM or more, the input abundances can be recovered exactly. This agrees within $\pm0.1$FWHM with the value of 0.6 found in the previous theory section (\ref{blending}). At 1$\times$FWHM the separation of the lines is visible, and the abundances can be derived as if the lines were not blended.

In one test spectrum we introduced an extra line\footnote{0.15\,\AA~ from the Cr line} with an abundance of 0.1\,dex into the blend as an unknown blend. This increased the derived Cr abundance by 0.1\,dex. 

Finally, if we assume the abundance of one of the two blending lines to be known (e.g., Cr), based on measurements from other clean Cr lines, the abundance of the Al line can be perfectly recovered regardless of FWHM separation even when the centres of the two blending lines coincide.

The observed 4MOST spectra will not necessarily be shifted perfectly to rest wavelength, but the radial velocity might be off by $\sim1$\,\kms. According to another scientific requirement, the radial velocity must be accurate to within $\pm1-2$\,\kms. In metal-rich giants most of the lines will be very strong or even saturated, which means that the line profiles will no longer maintain a Gaussian shape\footnote{This can also afffect the accuracy of the radial velocity shift, since saturated lines tend to yield more uncertain shifts.}. Combining all these factors and furthermore adding the radial velocity as a free parameter will aggravate the abundance recovery from a blend containing a saturated and a weaker line. Thus, by considering a short wavelength range consisting mainly of non-Gaussian shaped strong lines, could lead to problems where an uncertain radial velocity no longer causes a simple systematic offset.

\begin{figure}[h!]
\centering
\resizebox{\hsize}{!}{\includegraphics{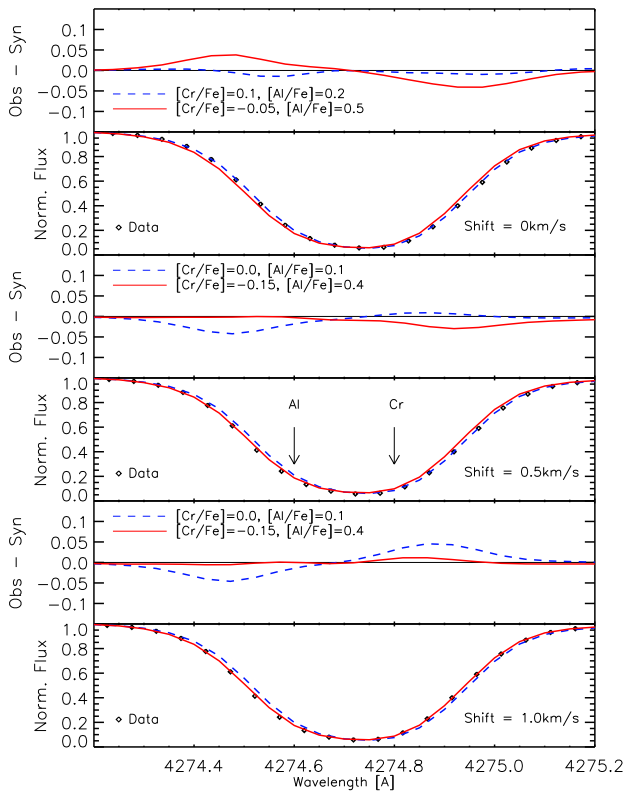}}
\caption{Mock data (observations -- black diamonds) with different radial velocity shifts (i.e., offset by 0, 0.5, 1\,\kms\, from top to bottom) compared to syntheses with different [Cr/Fe] and [Al/Fe] abundances. All spectra have been broadened to the 4MOST resolution. Residuals (observations - syntheses) show the quality of the fit. At an offset of 0.5\,\kms\, the syntheses are equally good/bad. In order to derive abundances that are exact within $\pm0.1$\,dex the radial velocity shift needs to be exact (0\,\kms) or the separation of the blending lines needs to be larger than $1\times$FWHM.
\label{fig:badblend}}
\end{figure}

3) Therefore a different test was carried out on a 'worst case scenario', namely a more metal-rich giant (4500K/2.0/$-1.0$/2\,\kms) with numerous blends in the blue region. We used 0.1\,dex lower abundances\footnote{i.e., [Cr/Fe]=0.0 and [Al/Fe]=0.1\,dex} for this test. 
The saturation plus offset radial velocity can hide the weaker line and return an abundance that is off by up to 0.3--0.4\,dex (see Fig.~\ref{fig:badblend}). To obtain the correct abundance a larger separation ($> 1\times$FWHM) is needed. 
The shift of the line centre (mock data offset by 0, 0.5, and 1\,\kms see Fig.~\ref{fig:badblend}) is barely visible in the broad line with an almost flat core of the saturated line. Only under
careful ($\chi^2$) inspection this will be detected. In an
automated code this could result in a false minimum, that would
overestimate the abundance of the weaker line by up to 0.3-0.4\,dex, while
the stronger line could be reproduced within $\sim \pm0.2$\,dex.

The last test (3) illustrates future limitations in the automated pipeline abundance analyses of 'worst case scenarios' which is needed for a survey like 4MOST.
Not surprisingly, the more metal-rich giants are the hardest cases from
which to recover the abundances for each of the lines. The lines are extremely strong (saturated or even damped) in metal-rich giants,
making the abundance recovery even harder. This is why the
blue range with all its blends can generally not be used for more metal-rich
disk and bulge stars. This does not mean that this range does not contain any useful lines. Section~\ref{wavelength} shows that there are several clean lines in metal-rich giants for which we can derive abundances within the adopted uncertainty ($\pm0.1$\,dex). Special attention needs to be paid to very metal-rich giants to ensure that the radial velocity shift is accurate since this could degrade the abundances derived from strong blended lines.

Such an extreme case would be very unlikely at a metallicity lower than [Fe/H]$<-1$, where there are fewer (saturated) lines, and if the lines are not studied in isolation, a `wrong' radial velocity ($\pm1$\,\kms) should reveal itself by offsetting the surrounding lines.

This shows that as long as we deal with two weak, blended lines, these can be used for precise abundance measurements ($\pm0.1$\,dex) in the 4MOST HRS spectra, provided the line centres are separated by $\sim0.5\times$FWHM if all other parameters and input (e.g., atomic data, stellar parameters, and radial velocity) are well constrained. If we know the abundance for one of the blending elements from other clean lines, we can recover the abundance of the other component perfectly even if the two lines overlap. On the other hand, if the radial velocity is uncertain and the line is saturated or contains unknown blends, we will need a larger separation ($>1\times$FWHM) to recover the abundances from the blending lines.

\section{Conclusions and outlook}
\label{conclusion}

The future survey instrument 4MOST will provide data that are crucial
for making advances in cosmology, active galactic nuclei, and the
formation of galaxies. For the latter, the Milky Way will be used
as a model organism.

4MOST will allow for studies on large scales to understand galaxy
formation as well as on small scales by tracing the origin of heavy
elements.  Both of these are among the open questions in
astrophysics to date, and they cannot be answered to the same extent
using green and red spectral studies only. In particular, the important science case of metal-poor stars cannot be studied and answered using
only redder spectra.

With the current design plan, the instrument will be able to
simultaneously observe up to 2400 objects in one pointing, using fibers
feeding high- and low-resolution spectrographs. To be able to measure
the abundances of at least the 20 elements that are most important for
constraining the chemical evolution of the Galaxy, the blue arm of the
4MOST high-resolution spectrograph must provide a resolving power of
$R\sim18\,000$--$20\,000$, and a sampling of 2.5--3 pixels per resolution
element.

The blue spectral range is very crowded and therefore many lines blend. Here we
have shown that according to theory, the line centres of two weak, 
blending lines need to be separated by $\sim 0.6\times$FWHM to make an abundace recovery within $\pm0.1$\,dex uncertainty. 
Our
subsequent tests on synthetic spectra showed that the separation needs to be $\ga0.5
\times$FWHM in order to recover accurate abundances (within
$\pm0.1$dex) from both components (unless we a priori know the abundance
of one of the elements).  
The precision of the radial velocity
determination plays an important role in how accurately we can recover
the abundances from two blended strong lines.

To date only three high-resolution instruments are capable of observing
all the way down to the atmospheric cut-off while most others stop $\sim
50-100$\,nm redward of this. However, the $300$--$400$\,nm region is
important when observing halo stars no matter if the science cases focus
on constraining nuclear formation processes of elements with Z$\ge38$ or searching for
the origin of carbon enhanced metal-poor stars, which could be some of
the first stars formed after the Population III stars. However, to unravel the
origin of stars, nuclear processes, or large scale formation scenarios
of the Milky Way, a large sample of stars is needed at
$\mathrm{[Fe/H]}<-3$. The efficiency of the fibres needs to improve to cover a wider wavelength range at higher resolution in the blue to make best use of these blue-most wavelength regions. Alternatively, a
stronger initiative in the community focusing on blue-to-near-UV science
is needed, and even more time needs to be allocated to high-resolution
slit spectroscopy covering a deeper follow-up of these blue wavelength
ranges.

\begin{acknowledgements}
We thank the referee for useful and detailed comments.
  C. J. Hansen was supported by a research grant (VKR023371) from VILLUM
  FONDATION and by Sonderforschungsbereich SFB 881 "The Milky Way System"
  (subproject A5) of the German Research Foundation (DFG).  AK
  acknowledges the DFG for funding from Emmy-Noether grant Ko 4161/1.
  CJH is grateful to Olga Bellido for her useful comments and input to the paper. L. S. acknowledges the support of Chile’s Ministry of Economy, Development, and Tourism's Millennium Science Initiative through grant IC120009, awarded to The Millennium Institute of Astrophysics, MAS.
  EC is grateful to the FONDATION MERAC for funding her fellowship.
\end{acknowledgements}

 \bibliographystyle{an}

\begin{thebibliography}{}
\bibitem[{{Ad{\'e}n} {et~al.}(2011){Ad{\'e}n}, {Eriksson}, {Feltzing},
  {Grebel}, {Koch}, \& {Wilkinson}}]{Aden2011}
{Ad{\'e}n}, D., {Eriksson}, K., {Feltzing}, S., {et~al.} 2011, \aap, 525, A153

\bibitem[{{Aoki} {et~al.}(2007)}]{Aoki2007} {Aoki}, W., {Beers}, T.~C., 
{Christlieb}, N., {et al.} 2007, \apj, 655, 492 

\bibitem[{{Beers} \& {Christlieb}(2005)}]{Beers2005}
{Beers}, T.~C. \& {Christlieb}, N. 2005, \araa, 43, 531

\bibitem[{{Bisterzo} {et~al.}(2012){Bisterzo}, {Gallino}, {Straniero},
  {Cristallo}, \& {K{\"a}ppeler}}]{Bisterzo2012}
{Bisterzo}, S., {Gallino}, R., {Straniero}, O., {Cristallo}, S., \&
  {K{\"a}ppeler}, F. 2012, \mnras, 422, 849

\bibitem[Bonifacio {et~al.}(2015){Bonifacio}, {Caffau}, {Spite}, {et al.},]{Bonifacio2015}{Bonifacio}, P., {Caffau}, E., {Spite}, M., {et al.} 2015, \aap, 579, A28 

  
\bibitem[{{Caffau} {et~al.}(2013){Caffau}, {Koch}, {Sbordone}, {Sartoretti},
  {Hansen}, {Royer}, {Leclerc}, {Bonifacio}, {Christlieb}, {Ludwig}, {Grebel},
  {de Jong}, {Chiappini}, {Walcher}, {Mignot}, {Feltzing}, {Cohen}, {Minchev},
  {Helmi}, {Piffl}, {Depagne}, \& {Schnurr}}]{Caffau2013}
{Caffau}, E., {Koch}, A., {Sbordone}, L., {et~al.} 2013, Astronomische
  Nachrichten, 334, 197

\bibitem[{{Carollo} {et~al.}(2007){Carollo}, {Beers}, {Lee}, {Chiba}, {Norris},
  {Wilhelm}, {Sivarani}, {Marsteller}, {Munn}, {Bailer-Jones}, {Fiorentin}, \&
  {York}}]{Carollo2007}
{Carollo}, D., {Beers}, T.~C., {Lee}, Y.~S., {et~al.} 2007, \nat, 450, 1020

\bibitem[{{Cayrel}(1988)}]{Cayrel1988}
{Cayrel}, R. 1988, in IAU Symposium, Vol. 132, The Impact of Very High S/N
  Spectroscopy on Stellar Physics, ed. G.~{Cayrel de Strobel} \& M.~{Spite},
  345

\bibitem[{{Cayrel} {et~al.}(2004){Cayrel}, {Depagne}, {Spite}, {Hill}, {Spite},
  {Fran{\c c}ois}, {Plez}, {Beers}, {Primas}, {Andersen}, {Barbuy},
  {Bonifacio}, {Molaro}, \& {Nordstr{\"o}m}}]{Cayrel2004}
{Cayrel}, R., {Depagne}, E., {Spite}, M., {et~al.} 2004, \aap, 416, 1117

\bibitem[{{Christlieb} {et~al.}(2004){Christlieb}, {Gustafsson}, {Korn},
  {Barklem}, {Beers}, {Bessell}, {Karlsson}, \&
  {Mizuno-Wiedner}}]{christlieb2004}
{Christlieb}, N., {Gustafsson}, B., {Korn}, A.~J., {et~al.} 2004, \apj, 603,
  708

\bibitem[{{Clarkson} {et~al.}(2008){Clarkson}, {Sahu}, {Anderson}, {Smith},
  {Brown}, {Rich}, {Casertano}, {Bond}, {Livio}, {Minniti}, {Panagia},
  {Renzini}, {Valenti}, \& {Zoccali}}]{Clarkson2008}
{Clarkson}, W., {Sahu}, K., {Anderson}, J., {et~al.} 2008, \apj, 684, 1110

\bibitem[{{Cowan} {et~al.}(2002){Cowan}, {Sneden}, {Burles}, {Ivans}, {Beers},
  {Truran}, {Lawler}, {Primas}, {Fuller}, {Pfeiffer}, \& {Kratz}}]{cowan2002}
{Cowan}, J.~J., {Sneden}, C., {Burles}, S., {et~al.} 2002, \apj, 572, 861

\bibitem[{{de Jong} {et~al.}(2014){de Jong}, {Barden}, {Bellido-Tirado},
  {Brynnel}, {Chiappini}, {Depagne}, {Haynes}, {Johl}, {Phillips}, {Schnurr},
  {Schwope}, {Walcher}, {Bauer}, {Cescutti}, {Cioni}, {Dionies}, {Enke},
  {Haynes}, {Kelz}, {Kitaura}, {Lamer}, {Minchev}, {M{\"u}ller}, {Nuza},
  {Olaya}, {Piffl}, {Popow}, {Saviauk}, {Steinmetz}, {Ural}, {Valentini},
  {Winkler}, {Wisotzki}, {Ansorge}, {Banerji}, {Gonzalez Solares}, {Irwin},
  {Kennicutt}, {King}, {McMahon}, {Koposov}, {Parry}, {Sun}, {Walton},
  {Finger}, {Iwert}, {Krumpe}, {Lizon}, {Mainieri}, {Amans}, {Bonifacio},
  {Cohen}, {Fran{\c c}ois}, {Jagourel}, {Mignot}, {Royer}, {Sartoretti},
  {Bender}, {Hess}, {Lang-Bardl}, {Muschielok}, {Schlichter}, {B{\"o}hringer},
  {Boller}, {Bongiorno}, {Brusa}, {Dwelly}, {Merloni}, {Nandra}, {Salvato},
  {Pragt}, {Navarro}, {Gerlofsma}, {Roelfsema}, {Dalton}, {Middleton}, {Tosh},
  {Boeche}, {Caffau}, {Christlieb}, {Grebel}, {Hansen}, {Koch}, {Ludwig},
  {Mandel}, {Quirrenbach}, {Sbordone}, {Seifert}, {Thimm}, {Helmi}, {trager},
  {Bensby}, {Feltzing}, {Ruchti}, {Edvardsson}, {Korn}, {Lind}, {Boland},
  {Colless}, {Frost}, {Gilbert}, {Gillingham}, {Lawrence}, {Legg}, {Saunders},
  {Sheinis}, {Driver}, {Robotham}, {Bacon}, {Caillier}, {Kosmalski}, {Laurent},
  \& {Richard}}]{Jong2014}
{de Jong}, R.~S., {Barden}, S., {Bellido-Tirado}, O., {et~al.} 2014, in Society
  of Photo-Optical Instrumentation Engineers (SPIE) Conference Series, Vol.
  9147, Society of Photo-Optical Instrumentation Engineers (SPIE) Conference
  Series, 0

\bibitem[{{de Jong} {et~al.}(2012){de Jong}, {Bellido-Tirado}, {Chiappini},
  {Depagne}, {Haynes}, {Johl}, {Schnurr}, {Schwope}, {Walcher}, {Dionies},
  {Haynes}, {Kelz}, {Kitaura}, {Lamer}, {Minchev}, {M{\"u}ller}, {Nuza},
  {Olaya}, {Piffl}, {Popow}, {Steinmetz}, {Ural}, {Williams}, {Winkler},
  {Wisotzki}, {Ansorge}, {Banerji}, {Gonzalez Solares}, {Irwin}, {Kennicutt},
  {King}, {McMahon}, {Koposov}, {Parry}, {Sun}, {Walton}, {Finger}, {Iwert},
  {Krumpe}, {Lizon}, {Vincenzo}, {Amans}, {Bonifacio}, {Cohen}, {Francois},
  {Jagourel}, {Mignot}, {Royer}, {Sartoretti}, {Bender}, {Grupp}, {Hess},
  {Lang-Bardl}, {Muschielok}, {B{\"o}hringer}, {Boller}, {Bongiorno}, {Brusa},
  {Dwelly}, {Merloni}, {Nandra}, {Salvato}, {Pragt}, {Navarro}, {Gerlofsma},
  {Roelfsema}, {Dalton}, {Middleton}, {Tosh}, {Boeche}, {Caffau}, {Christlieb},
  {Grebel}, {Hansen}, {Koch}, {Ludwig}, {Quirrenbach}, {Sbordone}, {Seifert},
  {Thimm}, {Trifonov}, {Helmi}, {Trager}, {Feltzing}, {Korn}, \&
  {Boland}}]{Jong2012}
{de Jong}, R.~S., {Bellido-Tirado}, O., {Chiappini}, C., {et~al.} 2012, in
  Society of Photo-Optical Instrumentation Engineers (SPIE) Conference Series,
  Vol. 8446, Society of Photo-Optical Instrumentation Engineers (SPIE)
  Conference Series, 0

\bibitem[{{Frebel} {et~al.}(2007){Frebel}, {Christlieb}, {Norris}, {Thom},
  {Beers}, \& {Rhee}}]{Frebel2007}
{Frebel}, A., {Christlieb}, N., {Norris}, J.~E., {et~al.} 2007, \apjl, 660,
  L117

\bibitem[{{Gilmore} {et~al.}(2013){Gilmore}, {Norris}, {Monaco}, {Yong},
  {Wyse}, \& {Geisler}}]{Gilmore2013}
{Gilmore}, G., {Norris}, J.~E., {Monaco}, L., {et~al.} 2013, \apj, 763, 61

\bibitem[{{Gilmore} {et~al.}(2012){Gilmore}, {Randich}, {Asplund}, {Binney},
  {Bonifacio}, {Drew}, {Feltzing}, {Ferguson}, {Jeffries}, {Micela},
  {Negueruela}, {Prusti}, {Rix}, {Vallenari}, {Alfaro}, {Allende-Prieto},
  {Babusiaux}, {Bensby}, {Blomme}, {Bragaglia}, {Flaccomio}, {Fran{\c c}ois},
  {Irwin}, {Koposov}, {Korn}, {Lanzafame}, {Pancino}, {Paunzen},
  {Recio-Blanco}, {Sacco}, {Smiljanic}, {Van Eck}, \& {Walton}}]{Gerry}
{Gilmore}, G., {Randich}, S., {Asplund}, M., {et~al.} 2012, The Messenger, 147,
  25

\bibitem[{{Gustafsson} {et~al.}(2008){Gustafsson}, {Edvardsson}, {Eriksson},
  {J{\o}rgensen}, {Nordlund}, \& {Plez}}]{Gustafsson2008}
{Gustafsson}, B., {Edvardsson}, B., {Eriksson}, K., {et~al.} 2008, \aap, 486,
  951

\bibitem[{Hansen} {et~al.}(2014a){Hansen}{Andersen}{Christlieb}]{Hansen2014a} {Hansen}, C.~J., {Andersen}, A.~C., \& {Christlieb}, N.\ 2014a, \aap, 568, A47 

\bibitem[{{Hansen} {et~al.}(2014b){Hansen}, {Montes}, \& {Arcones}}]{Hansen2014b}
{Hansen}, C.~J., {Montes}, F., \& {Arcones}, A. 2014b, \apj, 797, 123

\bibitem[{{Hansen} {et~al.}(2012){Hansen}, {Primas}, {Hartman}, {Kratz},
  {Wanajo}, {Leibundgut}, {Farouqi}, {Hallmann}, {Christlieb}, \&
  {Nilsson}}]{Hansen2012}
{Hansen}, C.~J., {Primas}, F., {Hartman}, H., {et~al.} 2012, \aap, 545, A31

\bibitem[{Hansen \& Primas}(2011)]{Hansen2011} {Hansen}, C.~J., \& {Primas}, F.\ 2011, \aap, 525, L5 

\bibitem[{{Hansen} {et~al.}(2014c){Hansen}, {Hansen},
  {Christlieb}, {Yong}, {Bessell}, {Garc{\'{\i}}a P{\'e}rez}, {Beers},
  {Placco}, {Frebel}, {Norris}, \& {Asplund}}]{THansen} {Hansen}, T.~T., {Hansen}, C.~J., {Christlieb}, N., {et~al.} 2014c, \apj,
  787, 162

\bibitem[{{Howes} {et~al.}(2014){Howes}, {Asplund}, {Casey}, {Keller}, {Yong},
  {Gilmore}, {Lind}, {Worley}, {Bessell}, {Casagrande}, {Marino}, {Nataf},
  {Owen}, {Da Costa}, {Schmidt}, {Tisserand}, {Randich}, {Feltzing},
  {Vallenari}, {Allende Prieto}, {Bensby}, {Flaccomio}, {Korn}, {Pancino},
  {Recio-Blanco}, {Smiljanic}, {Bergemann}, {Costado}, {Damiani}, {Heiter},
  {Hill}, {Hourihane}, {Jofr{\'e}}, {Lardo}, {de Laverny}, {Magrini},
  {Maiorca}, {Masseron}, {Morbidelli}, {Sacco}, {Minniti}, \&
  {Zoccali}}]{bulge_paper}
{Howes}, L.~M., {Asplund}, M., {Casey}, A.~R., {et~al.} 2014, \mnras, 445, 4241

\bibitem[{{Keller} {et~al.}(2014){Keller}, {Bessell}, {Frebel}, {Casey},
  {Asplund}, {Jacobson}, {Lind}, {Norris}, {Yong}, {Heger}, {Magic}, {da
  Costa}, {Schmidt}, \& {Tisserand}}]{Keller2014}
{Keller}, S.~C., {Bessell}, M.~S., {Frebel}, A., {et~al.} 2014, \nat, 506, 463

\bibitem[{{Koch}(2009)}]{Koch2009}
{Koch}, A. 2009, Astronomische Nachrichten, 330, 675

\bibitem[{{Koch} {et~al.}(2008){Koch}, {McWilliam}, {Grebel}, {Zucker}, \&
  {Belokurov}}]{Koch2008}
{Koch}, A., {McWilliam}, A., {Grebel}, E.~K., {Zucker}, D.~B., \& {Belokurov},
  V. 2008, \apjl, 688, L13

\bibitem[{{Koch} \& {Rich}(2014)}]{Koch2014}
{Koch}, A. \& {Rich}, R.~M. 2014, \apj, 794, 89

\bibitem[{{Koposov} {et~al.}(2010){Koposov}, {Rix}, \& {Hogg}}]{Koposov2010}
{Koposov}, S.~E., {Rix}, H.-W., \& {Hogg}, D.~W. 2010, \apj, 712, 260

\bibitem[{{Kupka} {et~al.}(2000){Kupka}, {Ryabchikova}, {Piskunov}, {Stempels},
  \& {Weiss}}]{VALD2000}
{Kupka}, F.~G., {Ryabchikova}, T.~A., {Piskunov}, N.~E., {Stempels}, H.~C., \&
  {Weiss}, W.~W. 2000, Baltic Astronomy, 9, 590

\bibitem[{{Lee} {et~al.}(2013){Lee}, {Beers}, {Masseron}, {Plez}, {Rockosi},
  {Sobeck}, {Yanny}, {Lucatello}, {Sivarani}, {Placco}, \& {Carollo}}]{Lee2013}
{Lee}, Y.~S., {Beers}, T.~C., {Masseron}, T., {et~al.} 2013, \aj, 146, 132

\bibitem[{{Lindegren} \& {Feltzing}(2013)}]{Lindegren2013}
{Lindegren}, L. \& {Feltzing}, S. 2013, \aap, 553, A94

\bibitem[{{Lucatello} {et~al.}(2006){Lucatello}, {Beers}, {Christlieb},
  {Barklem}, {Rossi}, {Marsteller}, {Sivarani}, \& {Lee}}]{Lucatello2006}
{Lucatello}, S., {Beers}, T.~C., {Christlieb}, N., {et~al.} 2006, \apjl, 652,
  L37

\bibitem[{{Masseron} {et~al.}(2010){Masseron}, {Johnson}, {Plez}, {van Eck},
  {Primas}, {Goriely}, \& {Jorissen}}]{Masseron2010}
{Masseron}, T., {Johnson}, J.~A., {Plez}, B., {et~al.} 2010, \aap, 509, A93

\bibitem[{{McWilliam} {et~al.}(1995){McWilliam}, {Preston}, {Sneden}, \&
  {Shectman}}]{McWilliam1995}
{McWilliam}, A., {Preston}, G.~W., {Sneden}, C., \& {Shectman}, S. 1995, \aj,
  109, 2736

\bibitem[{{Nissen} \& {Schuster}(2010)}]{Nissen2010}
{Nissen}, P.~E. \& {Schuster}, W.~J. 2010, \aap, 511, L10

\bibitem[{{Norris} {et~al.}(2007){Norris}, {Christlieb}, {Korn}, {Eriksson},
  {Bessell}, {Beers}, {Wisotzki}, \& {Reimers}}]{Norris2007}
{Norris}, J.~E., {Christlieb}, N., {Korn}, A.~J., {et~al.} 2007, \apj, 670, 774

\bibitem[{{Placco} {et~al.}(2014) {Placco}, {Frebel}, {Beers}, \& {Stancliffe}}]{Placco2014} {Placco}, V.~M., {Frebel}, 
A., {Beers}, T.~C., \& {Stancliffe}, R.~J.\ 2014, \apj, 797, 21 


\bibitem[{{Ruchti} {et~al.}(2014){Ruchti}, {Read}, {Feltzing}, {Pipino}, \&
  {Bensby}}]{Ruchti2014}
{Ruchti}, G.~R., {Read}, J.~I., {Feltzing}, S., {Pipino}, A., \& {Bensby}, T.
  2014, \mnras, 444, 515

\bibitem[{{Sch{\"o}nrich} {et~al.}(2011){Sch{\"o}nrich}, {Asplund}, \&
  {Casagrande}}]{Schoenrich}
{Sch{\"o}nrich}, R., {Asplund}, M., \& {Casagrande}, L. 2011, \mnras, 415, 3807

\bibitem[{{Searle} \& {Zinn}(1978)}]{Searle1978}
{Searle}, L. \& {Zinn}, R. 1978, \apj, 225, 357

\bibitem[{{Shetrone} {et~al.}(2003){Shetrone}, {Venn}, {Tolstoy}, {Primas},
  {Hill}, \& {Kaufer}}]{Shetrone2003}
{Shetrone}, M., {Venn}, K.~A., {Tolstoy}, E., {et~al.} 2003, \aj, 125, 684

\bibitem[{{Sneden} {et~al.}(2008){Sneden}, {Cowan}, \& {Gallino}}]{Sneden2008}
{Sneden}, C., {Cowan}, J.~J., \& {Gallino}, R. 2008, \araa, 46, 241

\bibitem[{{Sneden} {et~al.}(2003){Sneden}, {Cowan}, {Lawler}, {Ivans},
  {Burles}, {Beers}, {Primas}, {Hill}, {Truran}, {Fuller}, {Pfeiffer}, \&
  {Kratz}}]{Sneden2003}
{Sneden}, C., {Cowan}, J.~J., {Lawler}, J.~E., {et~al.} 2003, \apj, 591, 936

\bibitem[{{Sneden}(1973)}]{Sneden1973}
{Sneden}, C.~A. 1973, PhD thesis, The University Of Texas At Austin.

\bibitem[{{Venn} {et~al.}(2004){Venn}, {Irwin}, {Shetrone}, {Tout}, {Hill}, \&
  {Tolstoy}}]{Venn2004}
{Venn}, K.~A., {Irwin}, M., {Shetrone}, M.~D., {et~al.} 2004, \aj, 128, 1177

\bibitem[{{Williams} {et~al.}(2011){Williams}, {Steinmetz}, {Sharma},
  {Bland-Hawthorn}, {de Jong}, {Seabroke}, {Helmi}, {Freeman}, {Binney},
  {Minchev}, {Bienaym{\'e}}, {Campbell}, {Fulbright}, {Gibson}, {Gilmore},
  {Grebel}, {Munari}, {Navarro}, {Parker}, {Reid}, {Siebert}, {Siviero},
  {Watson}, {Wyse}, \& {Zwitter}}]{Williams2011}
{Williams}, M.~E.~K., {Steinmetz}, M., {Sharma}, S., {et~al.} 2011, \apj, 728,
  102

\end{thebibliography}

\end{document}